\documentclass[12pt]{article}

\usepackage{epsfig}

\newcommand{\g}{\gamma}

\newcommand{\si}{\sigma}

\newcommand{\simgt}{\,\rlap{\lower 3.5 pt \hbox{$\mathchar \sim$}} \raise 1pt
 \hbox {$>$}\,}
\newcommand{\simlt}{\,\rlap{\lower 3.5 pt \hbox{$\mathchar \sim$}} \raise 1pt
 \hbox {$<$}\,}
\newcommand{\equ}[2]{\begin{equation} \label{#1} #2 \end{equation} }

{\end{list}}
\newcounter{enumct}

\begin{document}
 
%set sloppy attitude to line breaks
\sloppy

%do not number pages
\pagestyle{empty}
\begin{center}

{\LARGE\bf Low and High $Q^2$ Jet Production }\\[4mm]
{\LARGE\bf in $\gamma^*\gamma$-Scattering
  at $e^+e^-$ Colliders }\\[5mm]
{\Large B.~P\"otter} \\[3mm]
{\it II. Institut f\"ur Theoretische Physik,\footnote{Supported by
    Bundesministerium f\"ur Forschung und Technologie, Bonn, Germany, 
    under Contract 05~7~HH~92P~(0), and by EU Fourth Framework Program
    {\it Training and Mobility of Researchers} through Network {\it
      Quantum Chromodynamics and Deep Structure of Elementary
      Particles} under Contract FMRX--CT98--0194 (DG12 MIHT).}
 Universit\"at Hamburg,}\\[1mm]
{\it  Luruper Chaussee 149, D-22761 Hamburg, Germany}\\[5mm] 

{\bf Abstract}\\[1mm]
\begin{minipage}[t]{140mm}
  I review recent results on NLO calculations in QCD for inclusive jet
  production in $\gamma^*\gamma$-scattering at $e^+e^-$ colliders. I
  discuss the region of large $Q^2$ (DIS regime) and the transition to
  small $Q^2$ (virtual photoproduction regime). The limit to real
  photoproduction is performed and compared with existing calculations.
\end{minipage}\\[5mm]

\rule{160mm}{0.4mm}

\end{center}

%*******************************************************************
\section{Introduction}

Recently, considerable progress has been made in investigating the 
structure of the virtual photon in jet production from $eP$ scattering
at HERA experimentally \cite{1,2,3,4,5} and on the theoretical side
calculations are available in LO \cite{6,7,8} and NLO \cite{9,9b,10,11}. 
It is well-known that the real photon can not only couple directly
to the charge of the bare quarks but can also fluctuate into a bound
state und thus serve as a source of quarks and gluons. This resolved
component is described by a parton distribution function (PDF) of the
real photon. The resolved component of the virtual photon deviates
from that of the real photon in that it has a $Q^2$-dependence. 
If $Q^2$ is large compared to the transverse jet energy $E_T$, the
resolved component of the virtual photon is believed to be negligable. 
Since so far only limited data exist on the structure of the virtual
photon \cite{pluto}, the modeling of the $Q^2$-behaviour of the
virtual photon PDF is still rather ambiguous. Two groups have provided
LO parametrizations of the virtual photon PDF that fit the data
\cite{pluto}, namely Gl\"uck, Reya and Stratmann \cite{grs} (GRS), and
Schuler and Sj\"ostrand \cite{sas} (SaS). The GRS group has also
calculated the virtual photon PDF's in NLO, but unfortunately no
parametrization is available. In the limiting case $Q^2\to 0$, the
virtual photon PDF's reproduce the real photon PDF's. 

An alternative way to test the structure functions of the virtual
photon is in the reaction $\gamma^*(Q^2)+\gamma (P^2=0)\to
\mbox{jets}+X$, which can be obtained at $e^+e^-$ colliders by
single-tag experiments. Presently, these kind of experiments are carried
out at LEP \cite{12}. The virtuality of the probing photon $Q^2$ has
to be sufficiently small in comparison with $E_T$ to allow for a
hadronic component in the virtual photon. The real photon with
virtuality $P^2\simeq 0$ has both a direct pointlike and a resolved hadronic
part. Jet production in $\gamma\gamma$-scattering with both photons
being on-shell has been studied for some time experimentally
\cite{15,16} and NLO calculations are available
\cite{17,18,19,20}. The comparison between theory and experiment is
rather satisfactory (see e.g.\ \cite{21}). Here I review the
calculations presented in \cite{9b,14}, which describe jet
production in $\gamma^*\gamma$-scattering for large and small $Q^2$ at
NLO QCD level.

The LO process contributing to $\gamma^*\gamma\to\mbox{jets}$ is given
by the direct (D) coupling of the photons to the charge of the bare
quarks, which leads to two final state jets with finite $E_T$. Both, 
the real and the virtual photon can be resolved for small $Q^2$. I
denote the component with a single-resolved real photon as SR, the
single-resolved virtual photon as SR$^*$ and the contribution with two
resolved photons as double resolved (DR). Taking into account both the
transverse and longitudinal polarizations of the virtual photon, the
cross section $d\sigma_{e^+e^-}$ for $e^+e^-$-scattering is
conveniently written as the convolution 
\begin{eqnarray} 
  \frac{d\sigma_{e^+e^-}}{dQ^2dy_ady_b} = \sum_{a,b} \int dx_adx_b
  F_{\gamma /e^-}(y_b) f_{b/\gamma}(x_b)
  \frac{\alpha}{2\pi Q^2} \left[ \frac{1+(1-y_a)^2}{y_a}
  f^U_{a/\gamma^*}(x_a)d\sigma_{ab} \right. \nonumber \\ 
  + \left. \frac{2(1-y_a)}{y_a} f^L_{a/\gamma^*}(x_a)d\sigma_{ab}
  \right]  \label{e+e-}
\end{eqnarray}
Here, the variables $y_a,y_b$ describe the momentum fraction of the
photons $a,b$ in the electron and $x_a,x_b$ describe the momentum
fraction of the partons in the photons $a,b$.
The PDF's of the real and the virtual photon are $f_{b/\gamma}(x_b)$ and 
$f^{U,L}_{a/\gamma^*}(x_a)$, respectively, where $U$ and $L$ denote
the unpolarized and longitudinally polarized photon contributions.
The direct photon interactions are included in formula
(\ref{e+e-}) through delta functions. For the direct virtual photon one has 
the relation $f^{U,L}_{a/\gamma^*}d\sigma_{ab} = \delta
(1-x_a)d\sigma^{U,L}_{\gamma^*b}$, whereas for the direct real photon
the relation is $f_{a/\gamma}d\sigma_{ab} = \delta
(1-x_b)d\sigma_{\gamma b}$, where $d\sigma_{ab}$ refers to the
partonic cross section. The function $F_{\gamma/e^-}(y_b)$ describes
the spectrum of the real photons emitted from the electron according
to the Weizs\"acker-Williams approximation \cite{wwill}.

In the following I will first present the NLO results for the region
$Q^2>E_T^2$ and then will describe the transition to $Q^2<E_T^2$.

%*******************************************************************
\section{Jet Production in the Deep-Inelastic Region}

The deep-inelastic region is characterized by $Q^2\gg E_T^2$, down to
a region, where $Q^2\simgt E_T^2$. In this region, the virtual photon
has only a point-like component and the photon PDF is
given by a delta function, so that eqn (\ref{e+e-}) reduces to
\begin{equation}
  \frac{d\sigma_{e^+e^-}}{dQ^2dy_ady_b} = \sum_{a,b} \int dx_b
  F_{\gamma /e^-}(y_b) f_{b/\gamma}(x_b) \frac{\alpha}{2\pi Q^2}
  \left[ \frac{1+(1-y_a)^2}{y_a} d\sigma^U_{\gamma^*b} +
  \frac{2(1-y_a)}{y_a} d\sigma^L_{\gamma^*b} \right]
\end{equation}
This means, only the D and SR components contribute to the cross
sections. The partonic cross sections in LO consist of two final state
particles. The NLO corrections consist of the virtual and real
corrections, which both exhibit characteristic divergencies, which can
be extracted by using e.g.\ dimensional regularisation. 
The sum of real and virtual corrections is finite after factorization
of singularities from the initial state. The real corrections for the
D subprocess have been calculated with the phase-space slicing method
in \cite{14} and the subprocesses for the SR component can be taken
from the literature on DIS $ep$-scattering (see e.g.\ \cite{graud}). 

I want to discuss some characteristic results for inclusive single-jet
production under kinematical conditions that will be encountered at
LEP2, where the photons are emitted by colliding electrons and
positrons, both having the energy of $E_e=83.25$ GeV. For the parton
densities of the photon I use the NLO parametrization of Gl\"uck, Reya
and Vogt \cite{grv}, transformed from the DIS$_\g$ to the
$\overline{\mbox{MS}}$ scheme with $N_F=5$ flavors. The 
renormalization and factorization scales are set equal to $Q$.
Jets are combined in the final state according to the Snowmass jet
definition \cite{snow}. The $E_T$ distributions of the single-jet
inclusive cross section  
\equ{}{ \frac{d\si^{1jet}}{dE_TdQ^2} = \int d\eta
  \frac{d\si^{1jet}}{dE_TdQ^2d\eta}  } 
for the $Q^2$-values $Q^2=10$ and $100$ GeV$^2$ are shown in Fig.\ 1 a and
b. The rapidity is integrated over the central region $|\eta |<2$ in the
hadronic c.m.s. The NLO distributions of the direct (dashed) and the
resolved (dotted) component of the cross section and the sum (full
line) of the two are plotted in the $E_T$-range $E_T \in [3,11]$ GeV.
\begin{figure}[ttt]
  \unitlength1mm
  \begin{picture}(122,55)
    \put(3,-61){\epsfig{file=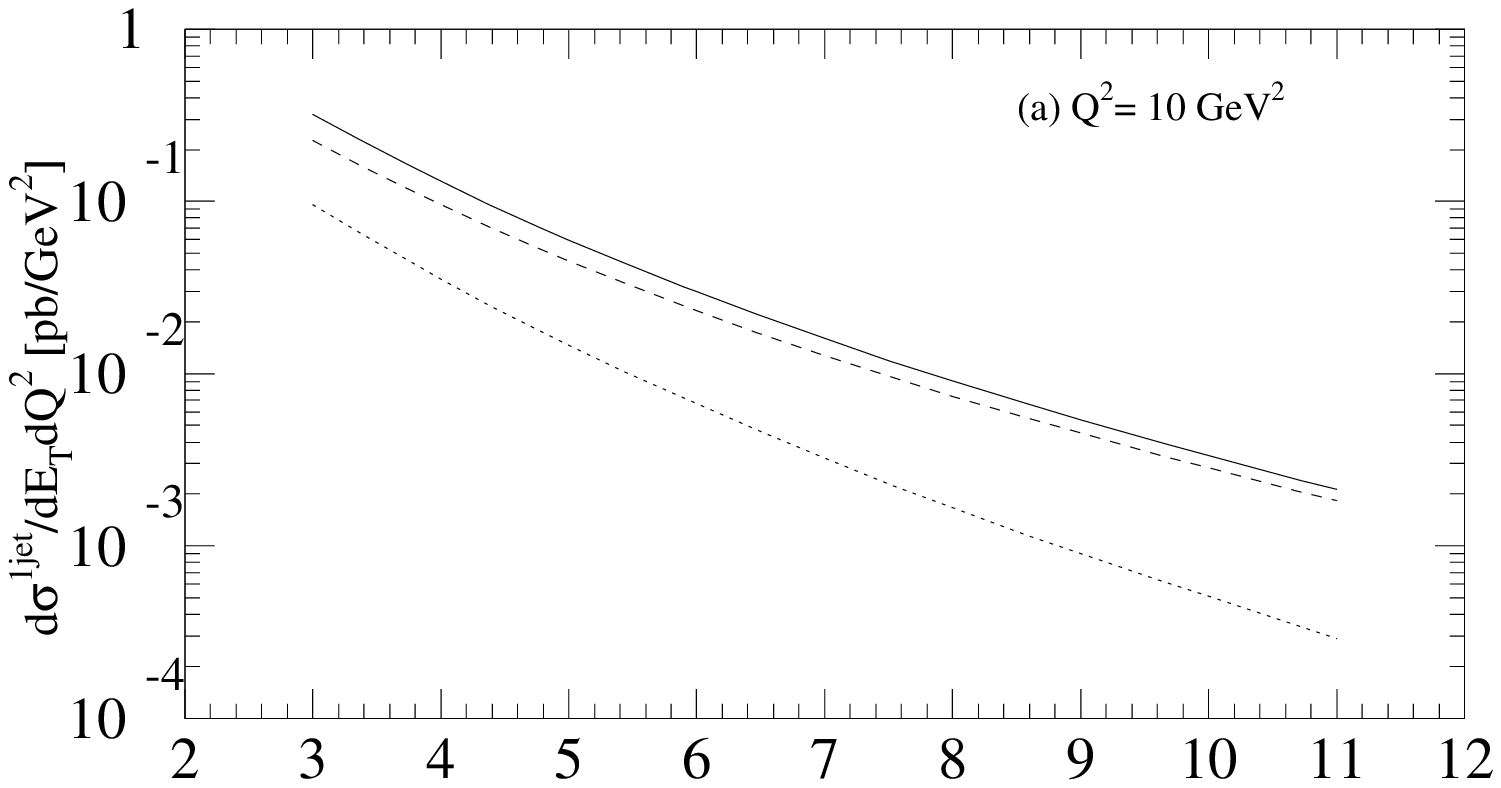,width=8.6cm,height=12.7cm}}
    \put(79,-61){\epsfig{file=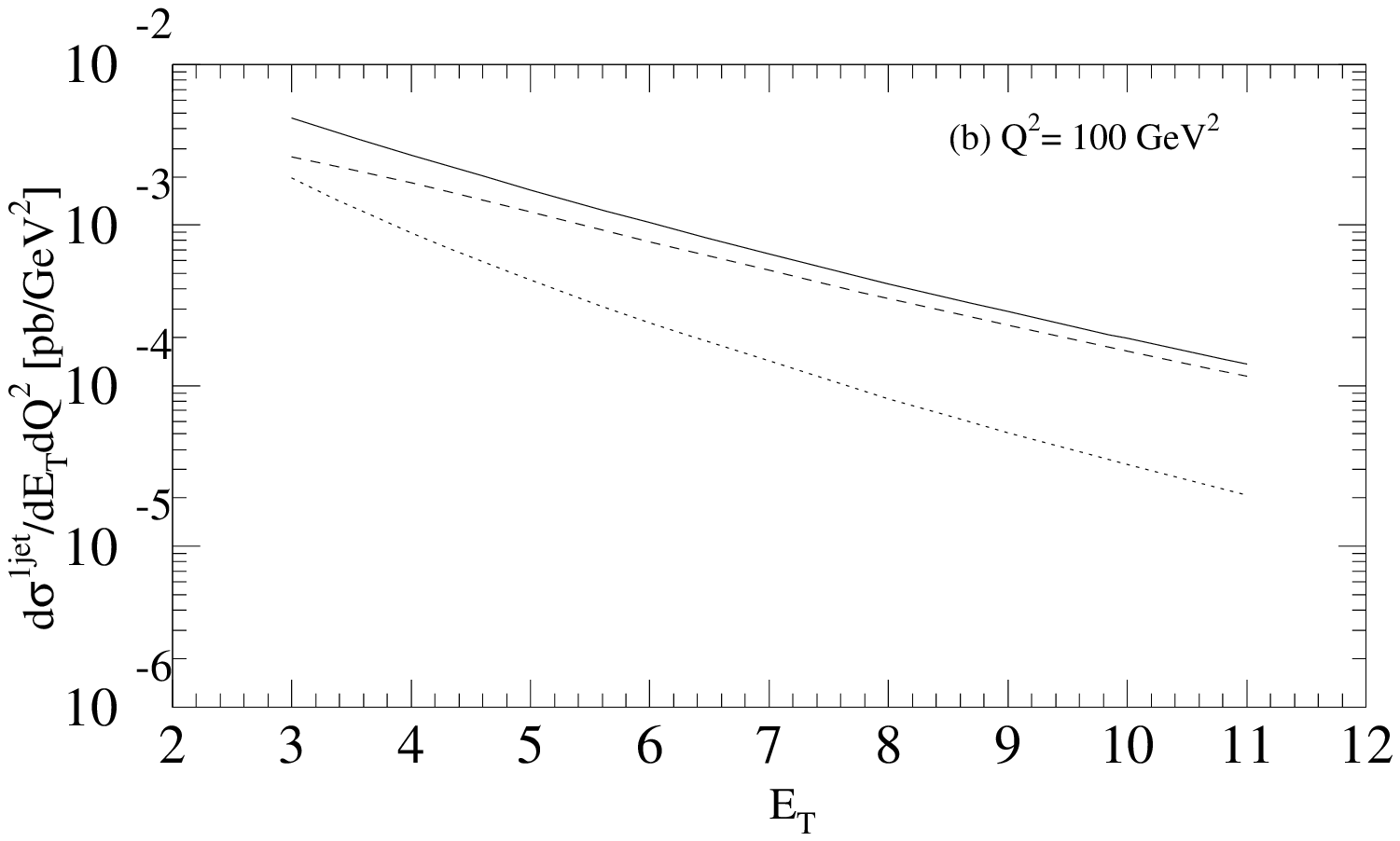,width=8.6cm,height=12.7cm}}
    \put(43,0){\footnotesize $E_T$}
  \end{picture}
  \caption{Inclusive single-jet 
        cross section $d\sigma^{1jet}/dE_TdQ^2$ integrated over $\eta$
        as a function of the transverse momentum $E_T$ for (a)
	$Q^2=10$ GeV$^2$ and (b) $Q^2=100$ GeV$^2$.}
\end{figure}

As can be seen, the direct component is the dominant one in the
whole $E_T$-range for the smaller $Q^2$-value. In addition the
resolved contribution falls off stronger with rising $E_T$ than the
direct component. The stronger fall-off of the resolved component holds
for the large $Q^2$-value as well, but here the resolved
component becomes nearly comparable to the direct contribution for the
smallest $E_T$ values. This is due to kinematical effect and stems
from the fact that the $\eta$-integration is restricted to $|\eta |<2$
(see \cite{14} for details). Since the resolved contribution is small
compared to the direct contribution, there is little hope
to learn about the parton distributions, especially the gluon
distributions, in the real photon from jet production in deep-inelastic
$e\g$-scattering. The fact that the resolved component is largely
suppressed for larger transverse energies $E_T$ agrees with the
expectation that the point-like part of the real photon is dominant
at large scales.

%*******************************************************************
\section{Transition to Real Photons}

Now I proceed to the case, where the virtuality $Q^2$ is small
compared to the transverse jet energy $E_T$. In this case, 
the $\gamma^*\to q\bar{q}$ splitting produces in the
phase-space-slicing method, in the limit of
the $q\bar{q}$-pair being collinear, the logarithm \cite{9,9b}
\begin{equation}
  M = \ln\left( 1 + \frac{y_ss}{zQ^2}\right) P_{q\leftarrow\gamma}(z)
  \ ,
\end{equation}
where $y_s$ is the phase-space-slicing parameter and $s$ is the
partonic cms energy, which is of the order of $E_T$. The logarithm
can be factorized from the cross section. The term becomes singular for
$Q^2=0$, which reproduces the real photon case, and then
has to be absorbed into the real photon PDF. This case is also
encountered on the real photon 
side in the DIS region and is described in \cite{14}. For finite $Q^2$
the logarithm is still large for $Q^2\ll E_T^2$, which suggests 
to absorbe a term proportional to $\ln(M_{\g }^2/Q^2)$ in the PDF of the
virtual photon. This is done in such a way that the
$\overline{\mbox{MS}}$ factorization result of the real photon is
obtained in the limit $Q^2\to 0$ \cite{9,9b}. When the leading
logarithmic term is subtracted from the D and SR components, a
resolved virtual photon component 
has to be added to the cross sections, which leads to the SR$^*$ and
DR contributions. The DR subprocesses can be taken from \cite{19,20}
and the SR$^*$ subprocesses follow from the SR ones for $Q^2\to 0$. 
The different subprocesses D, SR, SR$^*$ and DR have
been implemented  into the computer program {\tt JetViP} \cite{11}. 

In the following I present some numerical results for the LEP
conditions descibed above in the DIS case. Instead of the GRV PDF's of
the real photon, I consider only the $\overline{\mbox{MS}}$-GRS
\cite{grs} parametrization of the photon PDF here for real and virtual
photons. The number of flavors is set to $N_f=4$, adding the 
contributions from photon-gluon fusion by fixed order perturbation
theory. The renormalization and factorization scales are set equal to
$E_T$, in contrast to the DIS case above, so $\mu_R=M_\gamma
=M_{\gamma^*}=E_T$.

\begin{figure}[bbb]
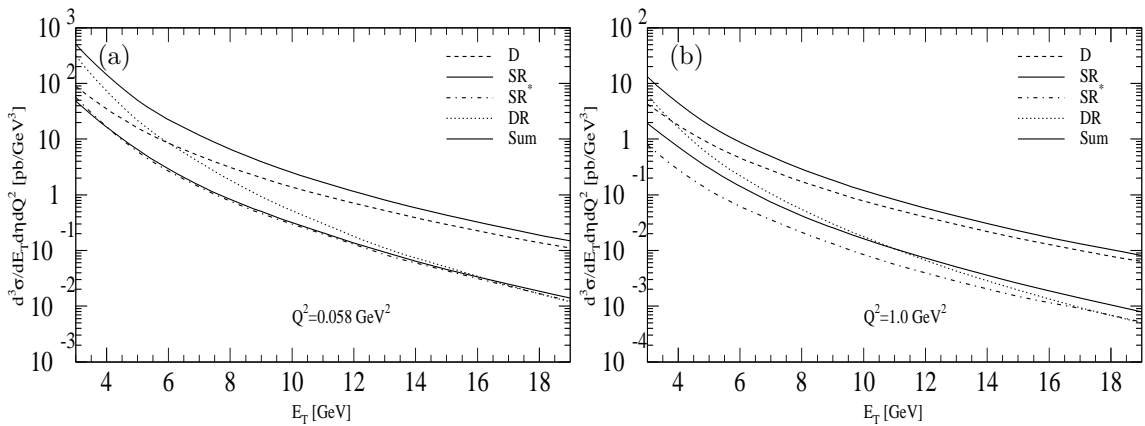

  \unitlength1mm
  \begin{picture}(122,55)
    \put(3,-61){\epsfig{file=plots/fig.10a,width=8.6cm,height=12.7cm}}
    \put(79,-61){\epsfig{file=plots/fig.10c,width=8.6cm,height=12.7cm}}
    \put(16,48.5){\footnotesize (a)}
    \put(92,48.5){\footnotesize (b)}
  \end{picture}
  \caption{Single-jet inclusive cross section integrated over $\eta
    \in [-2,2]$. The upper full curve is the sum of the D,
    SR, SR$^*$ and the DR components. (a) $Q^2=0.058$ GeV$^2$; 
    (b) $Q^2=1.0$ GeV$^2$.}
  \label{10}
\end{figure}

In Fig.~\ref{10} a and b the $E_T$ spectra for the virtualities
$Q^2=0.058$ and $1.0$ GeV$^2$ for the cross section
$d^3\sigma /dE_Td\eta dQ^2$ are shown, integrated over the interval
$-2\le \eta \le 2$. The value $Q^2_{eff}=0.058$ GeV$^2$
is chosen as to reproduce the $Q^2\simeq 0$ case. The
SR (lower full) and SR$^*$ (dash-dotted) curves coincide in Fig.\
\ref{10} a, where the real photon is approximated by the integrated
Weizs\"acker-Williams formula and the virtual photon has the value
of $Q^2_{eff}$. The full cross section (upper full curve) is dominated
by the DR component in the small $E_T$ range for the small $Q^2$
value. For $Q^2=1.0$ GeV$^2$, the DR and D contributions are of the
same order around $E_T=4$ GeV, but the DR component falls off quickly
for the higher $E_T$'s, leaving the D component as the dominant
contribution. This is expected, since the point-like coupling of the
photons is more important for larger $E_T$ and $Q^2$, as in the DIS
case. Since the virtual photon contribution is suppressed for larger
$Q^2$ the SR$^*$ contribution falls below  the SR curve when going to
higher values of $Q^2$. In all curves, both SR contributions do not
play an important role for the full cross section. Of course, all
contributions decrease with increasing $Q^2$.

\begin{figure}[ttt]
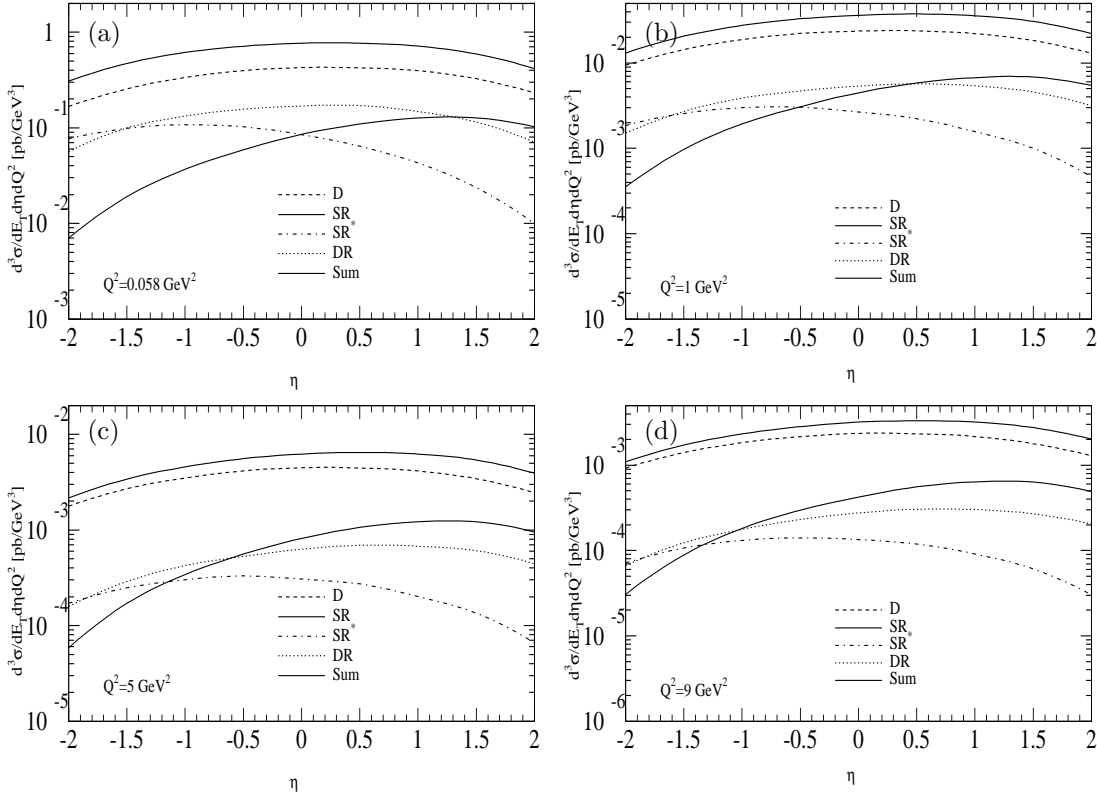

  \unitlength1mm
  \begin{picture}(122,100)
    \put(4,-5){\epsfig{file=plots/fig.11a,width=8.1cm,height=12cm}}
    \put(78,-5){\epsfig{file=plots/fig.11b,width=8.1cm,height=12cm}}
    \put(4,-58.5){\epsfig{file=plots/fig.11c,width=8.1cm,height=12cm}}
    \put(78,-58.5){\epsfig{file=plots/fig.11d,width=8.1cm,height=12cm}}
    \put(16,98){\footnotesize (a)}
    \put(90,98){\footnotesize (b)}
    \put(16,45){\footnotesize (c)}
    \put(90,45){\footnotesize (d)}
  \end{picture}
  \caption{Single-jet inclusive cross section as a function of $\eta$
    for fixed $E_T=10$ GeV. The upper full curve is
    the sum of the D, SR, SR$^*$ and the DR components.
    (a) $Q^2=0.058$ GeV$^2$; (b) $Q^2=1.0$ GeV$^2$; (c) $Q^2=5.0$
    GeV$^2$; (d) $Q^2=9.0$ GeV$^2$.}
  \label{11}
\end{figure}
 
I turn to the $\eta$-distribution of the single-jet cross section for
fixed $E_T=10$ GeV between $-2\le \eta \le 2$ for the virtualities
$Q^2=0.058, 1, 5$ and $9$ GeV$^2$, plotted in Fig.~\ref{11} a--d.
The D and DR distributions for the lowest virtuality
$Q^2_{eff}$ are almost symmetric, because of the identical energies of
the incoming leptons. The SR curve falls off for negative $\eta$,
whereas the SR$^*$ component is suppressed for positive $\eta$. Going
to higher $Q^2$ values, the D contribution stays more or less
symmetric and dominates the full cross section, as we have already
seen in Fig.~\ref{10} for the larger $E_T$ values. The components
containing contributions from the resolved virtual photon DR and
SR$^*$ fall of in the region of negative $\eta$ so that they become
more and more asymmetric. This is clear,  since we have chosen the
virtual photon to be incoming from the positive $z$-direction and the
resolved virtual photon contribution is decreasing for higher
virtualities. The DR and SR contributions are of the same magnitude in
the negative $\eta$ region and the DR component is dominant for the
larger $\eta$ values, where the resolved photon is more important. The
same holds for the D and SR$^*$ distributions in the negative $\eta$
region, only here the D component is far more dominant than the SR$^*$
component in the whole $\eta$ region. In conclusion it is clear that
all four components, especially the D and DR components, are important
in the cross section and thus, in addition to perturbative QCD, the
virtual photon PDF can be testet in low $Q^2$ jet production from
$\g^*\g$-scattering.

Finally, as a cross check of the calculations implemented in the
program {\tt JetViP} \cite{11}, I show a comparison of the four
components D, SR, SR$^*$ and DR in the limit $Q^2\to 0$ with the
photoproduction calculations of Kleinwort and Kramer \cite{19,20,21}
in Fig.~4. For these curves, the $Q^2$ was integrated by using the
Weizs\"acker-Williams approximation \cite{wwill} with $Q^2_{max}=1$
GeV$^2$. On the left side of Fig.~4 the comparison is made for the
$d\si/E_T$ as a function of $E_T$, where the rapidity has been
integrated out in $\eta \in [-2,2]$. The dots are the point from
Kleinwort and Kramer, whereas the curves are the predictions from {\tt
JetViP}. One sees a perfect agreement. This very good agreement holds
also for the right of Fig.~4, where $d\si/d\eta$ is shown as a
function of $\eta$. The transverse energy has been integrated out with
$E_T>3$ GeV. These comparisons show, that the subtraction of the leading
logarithmic term has been performed in a consistent way.
\begin{figure}[ttt]
  \unitlength1mm
  \begin{picture}(122,60)
    \put(11,68){\epsfig{file=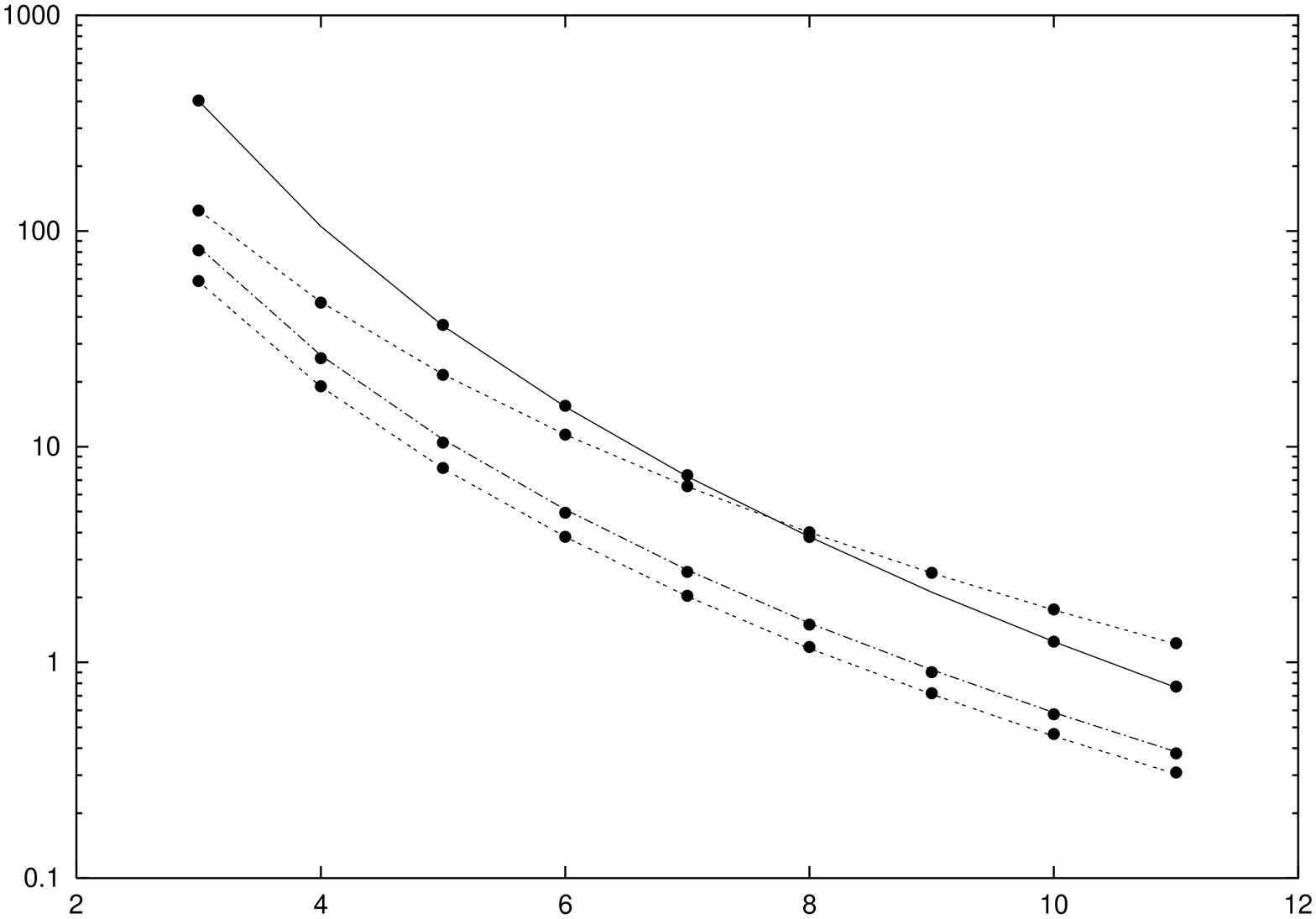,width=7.3cm,height=13.7cm,
        bbllx=6pt,bblly=100pt,bburx=647,bbury=1446pt,angle=270,clip=}}
    \put(89,68){\epsfig{file=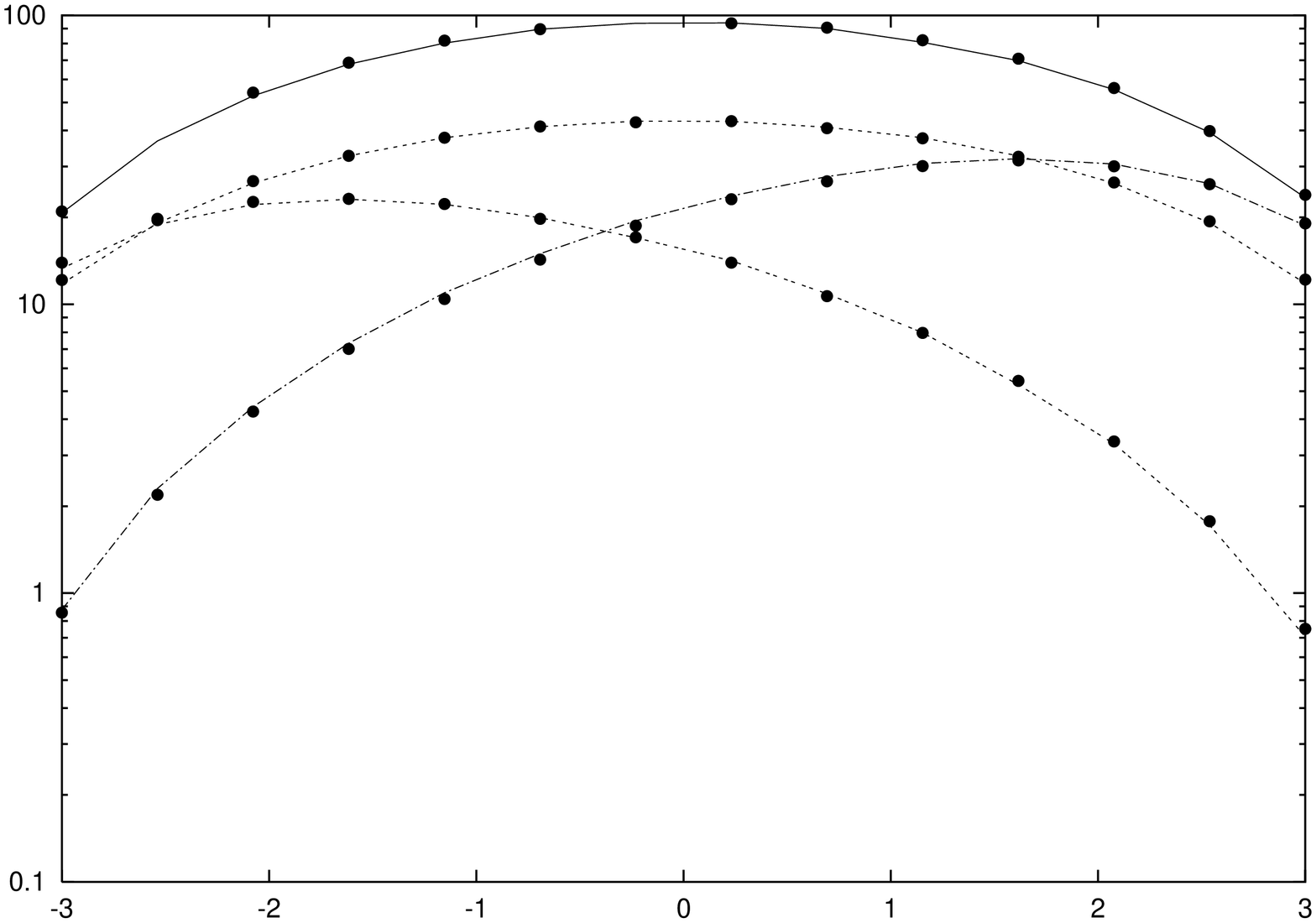,width=7.3cm,height=13.7cm,
        bbllx=0pt,bblly=90pt,bburx=647,bbury=1446pt,angle=270,clip=}}
    \put(42,0){\footnotesize $E_T$}
    \put(124,0){\footnotesize $\eta$}
    \put(2,35){$\frac{d\si}{dE_T}$}
    \put(82,35){$\frac{d\si}{d\eta}$}
  \end{picture}
  \caption{Single-jet inclusive cross section integrated over $\eta
    \in [-2,2]$ (left) and over $E_T>3$ GeV (right). The full line is
  the DR, the dotted is the D, the dashed is the SR and the
  dash-dotted the SR$^*$ component.}
\end{figure}
%

%*******************************************************************
\section{Conclusions}

I have reviewed results on NLO calculations in QCD for inclusive jet
production in $\gamma^*\gamma$-scattering at $e^+e^-$ colliders. 
I discussed the region of large $Q^2$ (DIS regime) and the transition to
small $Q^2$ (virtual photoproduction regime). The structure of the
virtual photon can be tested in this reaction, since the DR component
plays an important role in the cross sections. The limit to real
photoproduction was performed and the comparison with existing
calculations shows very good agreement.

%*********************************************************************

\end{document}